\documentclass[12pt,  amssymb]{article} 

\usepackage[dvips]{graphicx}            
\input epsf.sty
\usepackage{epsfig} 

\usepackage{amsfonts}
\usepackage{amssymb}

\input scrload

\newcommand{\newsection}[1]{
\addtocounter{section}{1} \setcounter{equation}{0}
\setcounter{subsection}{0} \addcontentsline{toc}{section}{\protect
\numberline{\arabic{section}}{{\rm #1}}} \vglue .6cm \pagebreak[3]
\noindent{\bf  \thesection. #1}\nopagebreak[4]\par\vskip .3cm}
\newcommand{\newsubsection}[1]{
\addtocounter{subsection}{1}
\addcontentsline{toc}{subsection}{\protect
\numberline{\arabic{section}.\arabic{subsection}}{#1}} \vglue .4cm
\pagebreak[3] \noindent{\it \thesubsection.
#1}\nopagebreak[4]\par\vskip .3cm}

\renewcommand{\theequation}{\thesection.\arabic{equation}}

\newlength{\extraspace}
\newlength{\extraspaces}
\newcounter{dummy}
\newcommand{\bc}{\begin{center}}
\newcommand{\ec}{\end{center}}
\newcommand{\be}{\begin{equation}
\addtolength{\abovedisplayskip}{\extraspaces}
\addtolength{\belowdisplayskip}{\extraspaces}
\addtolength{\abovedisplayshortskip}{\extraspace}
\addtolength{\belowdisplayshortskip}{\extraspace}}
\newcommand{\ee}{\end{equation}}

\newcommand{\ba}{\begin{eqnarray}
\addtolength{\abovedisplayskip}{\extraspaces}
\addtolength{\belowdisplayskip}{\extraspaces}
\addtolength{\abovedisplayshortskip}{\extraspace}
\addtolength{\belowdisplayshortskip}{\extraspace}}
\newcommand{\ea}{\end{eqnarray}}

\newcommand{\ban}{\begin{eqnarray*}
\addtolength{\abovedisplayskip}{\extraspaces}
\addtolength{\belowdisplayskip}{\extraspaces}
\addtolength{\abovedisplayshortskip}{\extraspace}
\addtolength{\belowdisplayshortskip}{\extraspace}}
\newcommand{\ean}{\end{eqnarray*}}
\newcommand{\baa}{
\addtocounter{equation}{1} \setcounter{dummy}{\value{equation}}
\setcounter{equation}{0}
\renewcommand{\theequation}{\thesection.\arabic{dummy}\alph{equation}}
\begin{eqnarray}
\addtolength{\abovedisplayskip}{\extraspaces}
\addtolength{\belowdisplayskip}{\extraspaces}
\addtolength{\abovedisplayshortskip}{\extraspace}
\addtolength{\belowdisplayshortskip}{\extraspace}}
\newcommand{\eaa}{
\end{eqnarray}
\setcounter{equation}{\value{dummy}}
\renewcommand{\theequation}{\thesection.\arabic{equation}}}

\newcommand{\bigo}{\mathcal{O}}
\newcommand{\bigO}{\mathcal{O}}
\newcommand{\bigL}{\mathcal{L}}
\newcommand{\bigD}{\mathcal{D}}

\newcommand{\tr}{\mbox{tr}}

\newcommand{\bK}{\mathcal{K}}

\newcommand{\besselK}{\mathcal{K}}
\newcommand{\myg}[2]{\hspace{.2cm}\mbox{\raisebox{#1}{\includegraphics{#2}}}\hspace{.2cm}}

\newcommand{\beq}{\begin{equation}}
\newcommand{\eeq}{\end{equation}}
\newcommand{\bea}{\begin{eqnarray}}
\newcommand{\eea}{\end{eqnarray}}

\newcommand{\ra}{\rangle}
\newcommand{\la}{\langle}

\begin{document}
\begin{flushright}
 February 2006 \\
 PUPT-2192\\
 YITP-SB-06-04\end{flushright}
\vspace{1.5cm}

\thispagestyle{empty}

\begin{center}
{\Large\bf Double-Trace Deformations, \\ Mixed Boundary Conditions and \\
Functional
Determinants 
 in AdS/CFT  \\[20mm] }

{\sc Thomas Hartman}\\[2.5mm]
{\it
Jefferson Physical Laboratory\\
Harvard University, Cambridge, MA 02138, USA}\\[5mm]

{\sc   Leonardo Rastelli}\\[2.5mm]
{\it
  C. N. Yang Institute for  Theoretical Physics\\
Stony Brook University \\
Stony Brook, NY 11794, USA}\\[15mm]

{\sc Abstract}

\end{center}

\noindent

According to the AdS/CFT dictionary,
 perturbing the large $N$ boundary     theory by
 a relevant double-trace deformation
of the form $f {\cal O}^2$
corresponds in the bulk
 to imposing ``mixed'' boundary conditions
 for the field
dual to ${\cal O}$.  
In this note we  address various aspects of this correspondence.
The change  $c_{UV} - c_{IR}$ of the central charge
between the UV ad IR fixed points is known from explicit calculations
(hep-th/0210093, hep-th/0212138) to be exactly the same  in the bulk
and in the boundary  theories. 
By comparing   the appropriate bulk and boundary
functional determinants, we give a simple ``kinematic'' 
explanation for this universal agreement. We also clarify the prescription for computing
AdS/CFT correlators with $\Delta_-$ boundary conditions.

  \vfill

\newpage

\renewcommand{\Large}{\normalsize}


\newsection{Introduction}

The purpose of this note is to give
a simple explanation for a successful
 prediction of the AdS/CFT correspondence,
 concerning the renormalization group flow triggered
 in a large $N$ CFT by  a  ``double-trace'' perturbation \cite{witten02, gubser02, gubser02b}.\footnote{
 Multitrace deformations were introduced in the context of AdS/CFT
 in \cite{ABS}. The correspondence between multitrace deformations of the
 boundary theory and boundary conditions in the bulk
 was understood in \cite{witten02, BSS} and further explored in {\it e.g.} \cite{Muck:2002gm, 
 Minces:2002wp, Petkou:2002bb, SS, Barbon:2002xk, gubser02, Klebanov:2002ja,   gubser02b,
 Girardello:2002pp, Nojiri:2003zq, Troost:2003ig, Strassler:2003ht, Barbon:2003uu, Minces:2004zr, Aharony:2005sh, Elitzur:2005kz}.}
A $d$-dimensional conformal field theory 
perturbed by a relevant deformation 
of the form $ f  \bigO^2$,
where $\bigO$ is a ``single-trace''  operator of dimension
$\Delta_- < d/2$,
flows in the IR to another conformal fixed point \cite{witten02}. At large $N$,
the IR theory is simply related to the UV theory -- by a Legendre
transformation with respect to the source for the operator  $\bigO$ \cite{gubser02b}.
In particular the conformal dimension
of $\bigO$ flows from $\Delta_-$ in the UV
to $\Delta_+ = d - \Delta_-$ in the IR. 
This phenomenon has a holographic counterpart in $(d+1)$-dimensional
Anti-de Sitter space, as
the flow between different
boundary conditions for the bulk field $\phi$ dual to $\bigO$ \cite{witten02}. The boundary
conditions preserve AdS isometries ({\it i.e.}, conformal
invariance) only at the extrema of the flow,
where they correspond to the two roots $\Delta_\pm$
in the usual AdS/CFT formula $\Delta (\Delta -d) = m^2$,
 $m$  being the mass of the scalar field $\phi$.
 The variation of the central
charge $c_{UV} - c_{IR}$ is a next-to-leading
effect (of order $O(1)$) 
in the large $N$ expansion. It  can be determined  in the boundary CFT by 
an explicit mode sum \cite{gubser02b}, 
and in the bulk  AdS theory by a one--loop evaluation of the effective potential
for $\phi$ \cite{gubser02}. The two calculations,
 while superficially quite different, are in perfect agreement.
 What is remarkable is the universality of this agreement,
 which holds for arbitrary spacetime dimension $d$, arbitrary
 conformal dimension $ \Delta_-$, and
independently  of  supersymmetry \cite{gubser02b}. This
 seems to call for a more  direct explanation.

Indeed, we find that the  bulk and boundary calculations
 can be cast in a way that makes their equivalence manifest.
 Both in the boundary and in the bulk, evaluation
of the central charge amounts to isolating the logarithmically
divergent term of the partition function -- 
 this is a UV divergence in the boundary, and an IR divergence
in the bulk, consistent with the familiar UV/IR connection.  The 
bulk and boundary partition functions are appropriate functional 
determinants, evaluated respectively in the space of bulk and 
boundary fields. In essence, we are able to write the bulk 
functional determinant in a way that makes it obviously identical 
to the boundary functional determinant. Our methods should be 
applicable to
 more general backgrounds, for example
AdS black hole backgrounds, where the bulk
partition function corresponds to the thermal partition function
of the boundary theory. They may also shed light on
 ``designer gravity''  \cite{designer}, where a similarly universal
 bulk/boundary correspondence should find a more intrinsic explanation. 
Our result is similar in spirit to the recent understanding of
 ``$\tau_{RR}$-minimization'' \cite{tau} as the boundary dual of the bulk procedure of ``$Z$-minimization'' \cite{Z}:
once the dust settles, the bulk and boundary calculations  are seen
to be isomorphic {\it step by step}. Of course, not all entries
 of the AdS/CFT dictionary are explained by such direct, purely 
 kinematic mechanisms. But when they are, it seems worthwhile to spell out these 
 mechanisms in detail.

We also take the opportunity  to clarify the prescription to compute
AdS/CFT correlators with the $\Delta_-$ choice of boundary conditions. 
In section 5 we explain in diagrammatic terms how the $\Delta_-$ boundary
conditions for the  AdS propagators lead automatically to 
the Legendre transform recipe postulated in \cite{witten99}. 
Finally, an appendix reviews the bulk calculation of Gubser and 
Mitra \cite{gubser02}, with a  slight modification of the 
regulator method, which allows us to check  the assumption in 
\cite{gubser02} that $c_{UV} = c_{IR}$ at the 
Breitenlohner-Freedman \cite{freedman82} bound.

\newsection{Boundary: RG flow triggered by a double-trace deformation}

To make this paper self-contained, we briefly review in
this section the field theoretic analysis of Gubser and Klebanov \cite{gubser02b}.
Consider the (Euclidean) partition function of a $d$-dimensional CFT,
 in the presence of the double-trace
deformation $\frac {f }{2} \bigO^2$, and with a source $J$ for the single-trace operator ${\cal O}$, 
\begin{equation} \label{Zf}
Z_f[J] = \left\langle \exp \left(-\int \frac{f}{2}\bigo^2 + \int J
\bigo\right)\right\rangle_0\,.
\end{equation}
Here $ \langle \, \dots \, \rangle_0$ denotes a correlator in
the unperturbed CFT. 
 In order for the deformation to be relevant, 
 the conformal dimension of ${\cal O}$ in the unperturbed theory is
 assumed to be
  $\Delta_-  \equiv \frac{d}{2} - \nu$,
with $\nu > 0$. (Unitarity implies the lower bound $\Delta_- \geq \frac{d}{2} -1$, or
$\nu \leq 1$.)   The deformation
parameter $f$ has engineering dimension $2\nu > 0$ and we expect
 that $f \to \infty$ in the IR. In fact,  
 the theory can be analyzed exactly in the large $N$ limit,
 all along the RG flow, using a 
 variant of the Hubbard-Stratonovich method. 
 This consists of introducing an auxiliary field $\sigma$, modifying the action as
\begin{equation}
S \rightarrow S - \frac{1}{2 f}\int (\sigma +
f\bigo)^2\,.
\end{equation}
The additional term does not change
 the physics -- the non-dynamical $\sigma$ field
can be integrated out giving back the original theory.
The partition function (\ref{Zf}) can be written as
\beq\label{eq:zaux}
Z_f[J] = \sqrt{\det (-f^{-1} {\bf 1})}\int\bigD\sigma
\left\langle\exp\int \left( \frac{\sigma^2}{2 f} +
(J+\sigma)\bigo\right)\right\rangle_0 \,.
\eeq
Dropping subleading terms in $1/N$,
\beq
\label{dropping}
\left\langle \exp \int (\sigma + J)\bigo\right\rangle_0 \cong
\exp\left(\frac{1}{2} \left\langle
\left(\int(\sigma+J)\bigo\right)^2\right\rangle_0\right)\,.
\eeq
To write a closed form expression
for $Z_f[J]$ 
it is convenient to
introduce the operator $\hat G$, defined in position space
as the convolution with the (undeformed)   two-point function of ${\cal O}$,
\begin{equation} \label{G}
(\hat G \sigma)(x) = \int \, d^dz\, 
\langle\bigo(x)\bigo(z)\rangle_0\, \sigma(z)= \int d^d z \,   \frac{\sigma(z)}{|x-z|^{2 \Delta_-} }\, ,
\end{equation}
and the operators
\begin{equation}
\hat K_f = 1 + f \hat G  \, , \hspace{1cm} \hat Q_f = \frac{\hat G}{1+ f
\hat G}\,.
\end{equation}
Clearly these kernels become diagonal in momentum space,
\begin{equation}\label{eq:flattwopoint}
G(k) = \int \frac{d^dk}{(2\pi)^d} \frac{e^{ikx}}{x^{2\Delta_-}} =
A_{\nu}k^{-2\nu}\,,     \quad A_\nu \equiv  2^{2 \nu} \pi^{d/2}\frac{\Gamma(\nu)}{\Gamma(\frac{d}{2} + \nu)}
\,.
\end{equation}
and
\beq
K_f = 1 +f  A_\nu k^{- 2 \nu} \, , \quad Q_f = \frac{A_\nu k^{-2 \nu}}{1 + f A_\nu k^{-2 \nu}}\,.
\eeq
With these notations in place,
the path integral over $\sigma$  gives
\beq\label{eq:uvpart}
Z_f[J] = \frac{1}{\sqrt{\det \hat K_f}}\exp\left(\int J \hat Q_f J\right) \,.
\eeq
Hence in the
presence of the deformation, the two point function is
\begin{equation}
\langle \bigO(x_1)\bigO(x_2)\rangle_f = \left.\frac{\delta^2 \log
Z_f[J]}{\delta J(x_1)\delta J(x_2)}\right|_{J=0} = Q_f(x_1,x_2)\,.
\end{equation}
In the IR, where $fG\gg 1$,  we have
\begin{equation}
Q_f(k) = \frac{1}{f} - \frac{1}{f^2 G(k)} + \cdots \, .
\end{equation}
In position space $Q_{f= \infty}(x,0) \sim 1/x^{2\Delta_+}$, with 
$\Delta_+ \equiv  \frac{d}{2} + \nu$.  The double-trace 
deformation triggers an RG flow that leads to a new IR fixed 
point;  the dimension of $\bigO$ flows from $\Delta_-$  in the UV 
to $\Delta_+$ in the IR. A simple manipulation of the generating 
functional (\ref{eq:zaux}, \ref{dropping}) shows that the UV and 
IR large $N$ CFTs are related by a Legendre transform 
\cite{gubser02b}. 

Still following \cite{gubser02b}, to determine the central
charge  we can place the CFT on a $d$-sphere of radius $R$ and 
compute
\be \label{c}
c \equiv \left \langle \int_{S^d_R}  d^dx \sqrt{g} \, T_\mu^\mu \right \rangle =
\frac{1}{d}R \frac{\partial}{\partial R}W[R] \, , \quad W[R] \equiv \log Z[J = 0, S^d_R] \, .
\ee
(Note that for $d=4$, this definition coincides -- up to a universal numerical factor -- with  
what is usually called the  ``$a$'' anomaly coefficient --  the coefficient
in front of the Euler density in the expression for the trace anomaly $T_\mu^\mu$
in a curved background. Indeed the  other curvature invariant 
is the Weyl tensor, which vanishes on the conformally flat ${\bf S}_R^d$.)
 As usual in quantum field theory, the partition function
is UV divergent. Introducing a UV (momentum) cutoff $\Lambda$, by 
scale invariance we must have $W = f(\Lambda R)
 = \alpha_d (\Lambda R)^d + \alpha_{d-1} (\Lambda R)^{d-1} + \cdots \alpha_0 \log(\Lambda R) 
 + \cdots$. Equation (\ref{c}) must be interpreted as valid for the renormalized
 $W[R]$ -- in practice this means that we focus on the logarithmic divergence,
 so that $c \equiv \alpha_0/d$.
 As  is  well-known, a non-zero anomaly
 can arise only for even
 $d$.

 We are interested in finding
\beq c_{IR} - c_{UV} =
\frac{1}{d}R \frac{\partial}{\partial R}(W_{f= \infty} [R] - W_{f=0}[R]) .\eeq
 From the analysis with the auxiliary field method,
 \beq\label{eq:cftans}
 W_{f_1}[R] - W_{f_2}[R] =
-\frac{1}{2}\tr\log\left( \frac{1+f_1 \hat G}{1 + f_2  \hat 
G}\right) =  -\frac{1}{2} \tr \log \left(\frac{\hat Q_{f_2}}{\hat 
Q_{f_1}} \right) \,.
\eeq
The eigenvalues $g_l$ of $\hat G$ on
${\bf S}_R^d$ are calculated using an expansion in spherical harmonics,
 \beq
G(x, x') =  \langle \bigO(x) \bigO(x')
\rangle_0 = \sum_{l,m} g_l Y_{lm}^*(x)Y_{lm}(x')\,,
 \eeq
which gives \cite{gubser02b}
\beq\label{eq:gl}
 g_l = R^{2\nu} \pi^{d/2}2^{2\nu}
\frac{\Gamma(\nu)\Gamma(l+\frac{d}{2}-\nu)}{\Gamma(\frac{d}{2}-\nu)\Gamma(l+\frac{d}{2}+\nu)}\,.
\eeq
Plugging this into (\ref{eq:cftans}), a fair amount of work still
needs to be done to extract the logarithmic divergence
$\sim \log(\Lambda R )$ and find the change in the central charge.
 The sum can be calculated \cite{gubser02b} with a zeta
function regulator, which automatically gets rid of the power law divergences
in $W$. In the following, we will 
directly compare the change in the CFT partition function (\ref{eq:cftans})
with the corresponding quantity in AdS space, so
we will not need to use anything beyond equations (\ref{eq:cftans}-\ref{eq:gl}).

\newsection{Bulk: mixed boundary conditions  }
\label{s:boundarycond}

If the unperturbed  CFT has a dual description as a gravitational
theory in  AdS$_{d+1}$, then its deformation by $\frac{f}{2} {\cal O}^2$
 has  a simple holographic interpretation \cite{witten02}.

Let us recall some basic facts about the AdS/CFT correspondence.
We write the Euclidean AdS metric as (the AdS scale is set to one)
\beq
ds^2 =\frac{1}{r^2} \left(dr^2 + A(r) g_{ij}(x) dx^i dx^j \right)\, ,
\eeq
where $x^i$, $i = 1, \cdots d$, parametrize the
conformal boundary at $r = 0$
and $A(r) \to 1$ as $ r \to 0$. Popular choices
 are Poincar\'e coordinates
$r \equiv x_0 $, $0 < x_0 < \infty$,
 $A = 1$, $g_{ij} = \delta_{ij}$,
for which the boundary is flat ${\mathbf R}^d$,
and hyperbolic coordinates $ r \equiv \rho$,
 $ 0 <  \rho \leq 2$, $A (\rho) =\left(\frac{4 - \rho^2}{4 }\right)^2 $, $g_{ij} dx^i dx^j  = d\Omega_d$,
for which the boundary is ${\bf S}^d$. For a free scalar field of 
mass $m$, the solution to the wave equation near the boundary $r 
\to 0$ takes the form
\beq \label{boundexp}
\phi(x, r) = r^{\Delta_+} [\alpha(x) + O(r^2) ] + 
r^{\Delta_-}[\beta(x) + O(r^2) ]  \,,
\eeq
where we have defined
\beq
\Delta_{\pm} = \frac{d}{2} \pm \nu \,,\hspace{1cm} \nu =
\sqrt{\frac{d^2}{4}+m^2}\,.
\eeq
The   Breitenlohner-Freedman bound \cite{freedman82} $m^2 \geq -\frac{d^2}{4}$
is necessary for stability and ensures that $\Delta_\pm$ are real.
For  general masses, the field must be quantized with the
boundary condition $\beta = 0$, the so-called
``regular'' choice of boundary conditions. 
In the range $-\frac{d^2}{4}  \leq m^2 < -\frac{d^2}{4}+1$  (equivalently $0 \leq \nu < 1$),
the ``irregular'' choice $\alpha = 0$ is also possible. This
can be understood (in Lorentzian signature)
from the fact that regular solutions are always normalizable,
while irregular solutions are normalizable only in this
restricted range of masses. 
With the regular choice of boundary conditions,
 we identify $\beta(x)$ with the source 
for an operator $\bigO$ in the boundary CFT  --
the  CFT action has a term $\int d^dx \, \beta(x) \, \bigO(x)$.
  Under the scaling isometry
$ r\rightarrow \lambda r$, $\phi$ is invariant hence
$\bigO$  has dimension $d-\Delta_- = \Delta_+$. 
The function $\alpha$ is identified with the expectation value 
of the operator, $\alpha = \la {\cal O} \ra$.
The irregular
choice of boundary conditions corresponds instead 
to identifying $\alpha(x)$ as the source of the boundary
operator and $\beta$ as the vev \cite{witten99}; then $\bigO$ has dimension $\Delta_-$.  
In going from the regular to the irregular boundary conditions, 
the roles of the source and of the vev are reversed,
and the generating functions of correlation functions 
of the two theories are related by a Legendre transformation \cite{witten99}. 
In section 5 we shall give a direct proof of this fact.

In fact in the mass range $-\frac{d^2}{4}  \leq m^2 < 
-\frac{d^2}{4}+1$, a large class of boundary conditions are 
possible, of the form $\alpha(x) = F[\beta(x)]$, where $F$ is any 
real functional. For generic $F$ the AdS isometry group is not 
preserved. Witten \cite{witten02} has interpreted these general boundary 
conditions as a multi--trace deformation of the boundary CFT, of 
the form $\int  d^d x \, {\cal W}[ {\cal O}(x)]$, with $F [\beta] 
= \frac{\delta {\cal W} }{\delta \beta}$. In the  case of a 
double-trace deformation, ${\cal W}({\cal O}) = \frac{f}{2} {\cal 
O}^2$, which leads to the boundary conditions\footnote{ Here 
$\tilde f   = {\rm const} \cdot f$ where the proportionality 
constant depends on the conventional normalization of ${\cal O}$ 
and will be determined below. In section 2, we normalized ${\cal 
O}$ to have unit two-point function in the unperturbed theory, 
see (\ref{G}). This normalization differs from the one obtained 
by taking functional derivatives with respect to the source 
$\alpha(x)$, which is the normalization implicit in Witten's 
prescription. This is why
  $f \to \tilde f$ in (\ref{mixed}).}
\beq \label{mixed}
\alpha(x) - \tilde f \beta(x) = 0 \,.
\eeq
This dovetails perfectly with the field theoretic analysis of the previous
section. In the UV, $f \to 0$ and the field is quantized with
irregular boundary conditions, so that the dual operator 
${\cal O}$ has $\Delta_-$ dimension; in the IR $ f \to \infty$
the regular boundary conditions are approached and ${\cal O}$ has dimension $\Delta_+$.

\newsubsection{A regulator}

To make precise sense of the AdS/CFT prescription it is often 
necessary (see {\it e.g.} \cite{ras98, skenderis98}) to introduce 
an IR bulk regulator -- which by the UV/IR connection plays the 
role of a boundary UV regulator. We cut off the infinite volume 
of AdS space by restricting the radial coordinate to $r \geq 
\epsilon > 0$, and setup the boundary value problem on the 
surface $r = \epsilon$.

Let us see how to write the boundary conditions (\ref{mixed}) in 
the presence of this cutoff.\footnote{See also the closely related 
discussion in \cite{BSS, SS}.} The action for a scalar field is
\beq
S = \int d^dx dr \sqrt{g} \left(\frac{1}{2}(\partial\phi)^2
 +\frac{1}{2} m^2\phi^2\right) + \int d^dx\sqrt{g}\bigL_{bdry}|_{r =
\epsilon}\,.
\eeq
Varying the field, we find upon integrating by parts and
using the bulk equations of motion,
\beq
\delta S = \int d^d x \sqrt{g}\, \left[\delta \phi \, \partial \phi
\cdot \hat{n} +\delta \bigL_{bdry}\right]_{r=\epsilon} \, ,
\eeq
where $\hat n =   \epsilon \hat r$ is the unit vector specifying the
normal to the boundary. In order for the variational problem to be well-defined,
 we must choose the boundary condition
for the field to cancel the contribution of the boundary term in
the action.  To reproduce the deformation $\frac{f}{2}\bigO^2$
added to the CFT, we choose the boundary term
\beq
\bigL_{bdry} = \frac{1}{2}\gamma \phi^2\,,
\eeq
which dictates the mixed Neumann/Dirichlet boundary
conditions\footnote{For all $\gamma$, Dirichlet conditions $\delta \phi = 0$
are also consistent, but we choose to impose (\ref{ND}).}
\beq \label{ND}
\gamma \, \phi(x, \epsilon) + \partial\phi(x, \epsilon)\cdot\hat{n} = 0\,.
\eeq
Now we must relate $f$ to $\gamma$.  To this end,
we are going to compute the two-point function using the AdS/CFT
prescription and compare it with the field theoretic
result of the previous section.  In AdS/CFT,
we are instructed to evaluate the on-shell
bulk action as a functional of the boundary source $\phi_b(x)$ \cite{gub98, wit98}.
For our mixed boundary conditions, the appropriate
boundary value problem is
\bea \label{boundaryvalue}
(\Box - m^2) \phi(x,r) & =&  0 \nonumber \\
\gamma \, \phi(x, \epsilon) + \partial \phi(x, \epsilon)\cdot\hat{n} & =&
\phi_b(x)\,  .
\eea
Plugging the solution of (\ref{boundaryvalue})
 back into the action and integrating by parts,
 \bea\label{eq:onshellaction}
S_{on-shell}[\phi_b] & =& \frac{1}{2}\int  d^dx \sqrt{g}\, [\phi(x ,r)\big(
\partial \phi(x,r)\cdot \hat{n} + \gamma \phi(x,r)\big]|_{r=\epsilon}  \nonumber \\ & = &
\frac{1}{2}\int  d^dx \sqrt{g}\, \phi(x ,\epsilon)\phi_b(x)\,. 
\eea To obtain an explicit solution, let us specialize to 
Poincar\'e coordinates, \be ds^2 = \frac{1}{x_0^2} (dx_0^2 + dx_i 
dx_i) \,  , \quad   x_0 \geq \epsilon \, , \ee and Fourier 
transform with respect to the flat boundary coordinates, $x_i \to 
k_i$, $i =1, \cdots d$, \be \phi(x_i, x_0) = \frac{1}{(2 
\pi)^{d/2}} \int d^dk  \, e^{i  k_i x_i} \, \phi(k_i, x_0) \,. 
\ee The wave equation reads \be \label{poincarelaplacian} \left[ 
-x_0^{d+1} \frac{\partial}{\partial x_0} \left( x_0^{-d +1}  
\frac{\partial}{\partial x_0} \right) + k^2 x_0^2 + m^2 \right] 
\phi(k, x_0)= 0 \ee with  $k = \sqrt{k_i k_i}$ . As is well-known 
(see {\it e.g.}  \cite{gub98,ras98}), the unique solution
 that is regular for $x_0 \to \infty$ is
\beq \label{psi}
\psi(k,x_0) \equiv x_0^{d/2}\besselK_\nu(kx_0) \, ,
\eeq
where $\besselK_\nu$ is the Bessel function.
Thus the solution  of the boundary value problem (\ref{boundaryvalue}) is
\beq \label{P}
\phi(k,r) = \left(\frac{\psi(k,r)}{\gamma\, \psi(k,\epsilon) +
\partial \psi(k,\epsilon)\cdot \hat{n}}\right) \phi_b(k)
\eeq
and the AdS/CFT dictionary 
 yields the CFT correlator
\bea\label{eq:cftcorrelator}
\langle \bigO(k) \bigO(k') \rangle &  = &  \left. -\frac{\delta^2 S_{on-shell}[\phi_b]}{\delta \phi_b(k) \delta \phi_b(k')}\right|_{\phi_b = 0} \\
&  = &-\epsilon^{-d}  \delta^d(k+k') \left(\frac{\psi(k,\epsilon)}{\gamma \, \psi(k,\epsilon)
+
\partial \psi(k,\epsilon)\cdot \hat{n}}\right)\,. \nonumber
\eea
Using the expansion of ${\cal K}_\nu (z) = z^{-\nu} [ 2^{-1+ \nu}\Gamma(\nu)+ O(z^2) ] + z^\nu [2^{-1-\nu} 
\Gamma(-\nu)+ O(z^2)]$,
we can extract the leading  behavior of the two point function
as $\epsilon \to 0$,
\beq
 \langle \bigO(k) \bigO(k')\rangle_{\gamma} =
  -\frac{ \epsilon^{-d} \delta^d(k+k')}
{ \gamma +\Delta_-  +  \left( \epsilon^{2\nu}
2^{-2\nu}\frac{\Gamma(-\nu)}{\Gamma(\nu)}(2\nu) \right) k^{2 \nu}+ O(\epsilon^2) }\,.
\eeq
Notice that in the mass range that we are considering, $\nu <1$ and the term $O(\epsilon^2)$ 
can indeed be neglected. On the other hand, the two-point function was computed directly
in the boundary CFT using the auxiliary field method,
\beq \label{directly}
\langle \bigO(k) \bigO(k')\rangle_{f} = \delta^d(k + k') Q_f (k) = \delta^d(k + k')\frac{ A_\nu}{  f A_\nu + k^{2 \nu} } \,.
\eeq
The two expressions coincide (up to an overall $k$-independent normalization that can be fixed by rescaling
the source $\phi_b(x)$), provided we identify
\beq\label{eq:fgamma}
\gamma = -\Delta_- - f \epsilon^{2\nu} \left( 2
\pi^{d/2}\frac{\Gamma(1-\nu)}{\Gamma\left(\Delta_-\right)}\right)\,.
\eeq
As $f \to \infty$, we find that $\gamma \to \infty$, and recover
the Dirichlet boundary value problem at $x_0 = \epsilon$, which 
is indeed the usual prescription for the ``regular''  $\Delta_+$ 
quantization. We see that the ``irregular'' $\Delta_-$ 
quantization $(f=0)$ corresponds to the specific choice $\gamma = 
- \Delta_-$  in the boundary conditions (\ref{mixed}). This can 
be understood by the following simple argument. The irregular 
boundary conditions consists of identifying the subleading term 
$\alpha(x)$ in the boundary expansion  of $\phi$ as the source 
for ${\cal O}$. Now
 the boundary source at $r = \epsilon$ is
 \bea \label{subleading}
 \phi_b(x) &= & \gamma \phi (x, \epsilon)  + \epsilon \partial_r \phi (x, r) |_{r = \epsilon}\\
 &=& 
 \epsilon^{\Delta_- } 
 [ (\gamma + \Delta_-) \beta(x) + O(\epsilon^2) ] + \epsilon^{\Delta_+}[ (\gamma + \Delta_+) \alpha(x) + O(\epsilon^2) ]  \, , \nonumber
  \eea
and indeed to cancel the dependence from the leading term $\beta$ 
we must choose $\gamma = - \Delta_-$.\footnote{Incidentally, we 
can also determine the relation between $f$ and the parameter 
$\tilde f$ introduced in (\ref{mixed}). Substituting 
(\ref{eq:fgamma}) in (\ref{subleading}) and setting $\phi_b=0$, 
$\alpha = \tilde f \beta$, we find $\tilde f = \frac{\pi^{d/2} 
\Gamma(1- \nu)}{\nu \Gamma(\Delta_-)}  f$, in agreement with 
\cite{gubser02b}.  We will not need this relation in the 
following.} In summary, the double-trace deformation $\int 
\frac{f}{2} {\cal O}^2$ of the boundary CFT corresponds to 
imposing mixed Neumann/Dirichlet boundary conditions (\ref{ND}) 
for the dual  scalar field $\phi$, with $\gamma$ and $f$ related 
by (\ref{eq:fgamma}). These boundary conditions become more
transparent in terms of the field $\chi$ introduced
in \cite{witten99}, locally related to $\phi$ as
\be
\phi(x_i, x_0) \equiv x_0^{\Delta_-} \chi(x_i, x_0)\,.
\ee
Then  $\Delta_-$ quantization corresponds
precisely to Neumann boundary conditions for $\chi$,
\be
\frac{\partial}{\partial {x_0}} \chi = 0 \, ,
\ee
while, of course, $\Delta_+$ quantization corresponds to  Dirichlet conditions.

General multitrace deformations
$\int d^dx {\cal W}[{\cal O}(x)]$ could be treated similarly,
by taking ${\cal L}_{bdry}={\cal W}(\phi)$.

\newsection{Relating the bulk and boundary partition functions}\label{ss:cftcentralcharge}

We are now going to show that the change in the partition
function $W$ induced by the double trace deformation is manifestly identical on
both sides of the correspondence.   In AdS, we look at the
variation of the partition function for a scalar
field of mass $m$ as we change boundary conditions,
\be
W_{\gamma_1}^{AdS} - W_{\gamma_2}^{AdS} = -\frac{1}{2} {\rm Tr}_{\gamma_1} \log (-\Box + m^2)
 +\frac{1}{2} {\rm Tr}_{\gamma_2} \log (-\Box + m^2)\,.
\ee
The trace  ``${\rm Tr}_\gamma$'' is over the bulk modes that obey
the  boundary conditions (\ref{eq:fgamma}). Separating the AdS coordinates
into a radial coordinate and $d$ coordinates parametrizing the boundary,
we can also write
\beq\label{eq:deltaVads}
W_{\gamma_1}^{AdS} - W_{\gamma_2}^{AdS} = -\frac{1}{2}\tr\log\left( \frac{\det_{\gamma_1}(-\Box
+ m^2)}{\det_{\gamma_2}(-\Box+m^2)}\right) \, ,
\eeq
where the trace ``${\rm tr}$'' is over the boundary modes
only, and the determinant ${\rm det}_\gamma$ is the product
of eigenvalues of the {\it radial}  wave-equation with $\gamma$ boundary conditions.
Here ``$(-\Box + m^2)$'' is a short-hand for the radial
differential operator, which depends on the eigenvalue
of the boundary Laplacian. For example, in Poincar\'e coordinates,
it is the second-order differential operator in $x_0$ given
in (\ref{poincarelaplacian}), which depends on the boundary
momentum $k$.

The corresponding quantity in the dual CFT is  (\ref{eq:cftans})
\beq
\label{eq:bdrypotential}
W^{CFT}_{f_1} - W^{CFT}_{f_2}   = -\frac{1}{2}\tr\log \left( \frac{1+f_1 G}{1 + f_2 G} \right)\,.
\eeq
Equating the change in the central charge
 on both sides of the AdS/CFT correspondence
thereby reduces to confirming that
\beq\label{eq:detratio}
\frac{\det_{\gamma_1}(-\Box+m^2)}{\det_{\gamma_2}(-\Box+m^2)} =
\frac{1+f_1 G}{1+f_2 G} = \frac{Q_{f_2}}{Q_{f_1}} \, .
\eeq
This will be our  task for the rest of the section.  

\newsubsection{One-dimensional determinants}\label{ss:colemantrick}
Fortunately, the ratio of one-dimensional determinants that
appears in (\ref{eq:detratio}) belongs to a well studied class of
problems. As a paradigmatic example, 
consider the
 two differential operators in one dimension
\begin{eqnarray}
{\cal Q} &=& -\frac{d^2}{dx^2} + R_{\cal Q}(x)\\
{\cal P} &=& -\frac{d^2}{dx^2} + R_{\cal P}(x) \, .
\end{eqnarray}
The problem is to find the ratio of functional
determinants
\begin{equation}
\frac{\det({\cal Q}-\lambda)}{\det({\cal P}-\lambda)} 
\, ,
\end{equation}
in the space of (square-integrable) functions ${\cal F} =  \{ f(x) \}$, $x \in [a , b]$, 
that satisfy specific boundary conditions -- for simplicity let us say Dirichlet
boundary conditions $f(a) = f(b) = 0$. This problem 
has an elegant solution.
 Define the functions
$q_\lambda$ and $p_\lambda$ to be eigenfunctions
of the two operators,
\begin{eqnarray}
\label{qpl}
{\cal Q} q_\lambda &=& \lambda q_\lambda\\
{\cal P} p_\lambda &=& \lambda p_\lambda \nonumber
\end{eqnarray}
with the boundary conditions 
\begin{eqnarray}
\label{qp}
q(b)=0  && \, \quad q'(b)=1 \\
 p(b) = 0 && \, \quad p'(b)=1\,. \nonumber 
\end{eqnarray}
Notice that in general $q_\lambda$ and $p_\lambda$
do not belong to ${\cal F}$. Then 
\begin{equation}\label{eq:detres}
\frac{\det({\cal Q}-\lambda)}{\det({\cal P}-\lambda)} =
\frac{q_\lambda(a)}{p_\lambda(a)}\,.
\end{equation}
In his famous lectures on instantons \cite{coleman}, Coleman gives the
following heuristic proof of this equation. Viewed as complex functions of
$\lambda$, both sides of the equation have the same poles and
zeros.  This is the case because for all $\lambda$, $q_\lambda$ satisfies 
one of the boundary conditions, $q(b)=0$; if it also
satisfies the other boundary condition
$q_\lambda(a)=0$, 
then $q_\lambda$ is an eigenfunction of ${\cal Q}$ in ${\cal F}$
and thus  ${\cal Q}-\lambda$ has vanishing determinant. The same applies to $p$
and ${\cal P}$.  Both sides of the equation also have the same limiting
behavior as $\lambda\rightarrow \infty$, so they must in fact be
the same function.

In \cite{Kirsten:2003py, kirsten04b},  a rigorous proof is given of a more 
general theorem. Formulas analogous to (\ref{eq:detres}) allow us 
to compute ratios of functional determinants
 for general Sturm-Liouville operators of the type
\begin{eqnarray}
{\cal Q}  & =&  -\frac{d}{dx}\left(S_{\cal Q}(x)\frac{d}{dx}\right) + R_{\cal Q}(x)\\
{\cal P} & =&  -\frac{d}{dx}\left(S_{\cal P}(x)\frac{d}{dx}\right) + R_{\cal P}(x)\, ,\nonumber
\end{eqnarray}
acting on the space
of functions  with some prescribed  
  mixed Neumann/Dirichlet boundary
conditions at the extrema. Consider for definiteness the space of 
functions ${\cal F}_\gamma$ with Dirichlet conditions at $x=b$ 
and mixed boundary conditions at $x=a$,
\begin{eqnarray*}
f(a) &=& 0\\
\gamma f(a) +  f'(a) &=& 0\,.
\end{eqnarray*}
Since the boundary conditions are parametrized by  $\gamma$, we
will denote the resulting determinant by $\det_\gamma$. If 
$q_\lambda$ and $p_\lambda$ are defined again as in (\ref{qpl}, 
\ref{qp}), then provided that $S_{\cal Q}=S_{\cal P}$ one has 
\cite{kirsten04b}
\beq\label{eq:colemantrick}
\frac{\det_\gamma {\cal Q}}{\det_\gamma {\cal P}} = \frac{\gamma q(a) +  q'(a) }{ \gamma p(a) + 
p'(a) }\, ,
\eeq
where  we have relabelled $q \equiv q_{\lambda =0}$, $p \equiv 
p_{\lambda= 0}$.

\begin{figure}[t]
\centering \epsfig{file=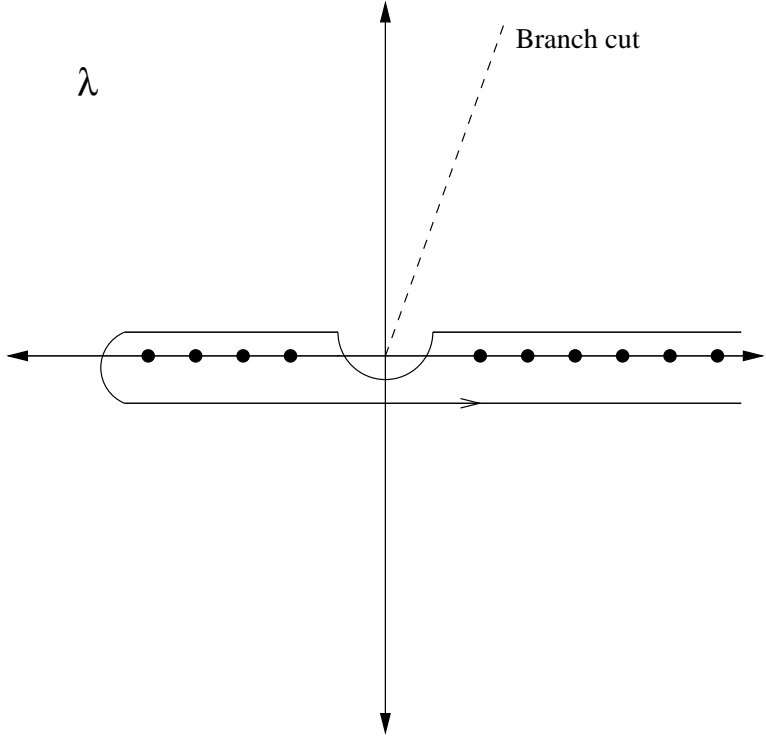, width=4in}
\caption{\textit{The contour and branch cut in Eq. (\ref{bobservation}) \cite{kirsten04b}.\label{fig:roots}}}
\end{figure}

One final complication is that in this formula, the same choice of
$\gamma$ must appear in both determinants, while the two operators ${\cal P}$
and ${\cal Q}$ are different. To check equation (\ref{eq:detratio}), we
need to compute a ratio of determinants where the operators are
the same
but the boundary conditions are different.  The obvious guess in 
this case is
\beq\label{eq:finalcoleman}
\frac{\det_{\gamma_1}{\cal  P}}{\det_{\gamma_2} {\cal P}} =\frac{ 
\gamma_1 p(a) +  p'(a) }{ \gamma_2 p(a) + p'(a) } \, ,
\eeq
which has the correct poles and zeros.
We will prove this formula by generalizing the proof given in  
\cite{kirsten04b}.  The zeta function of the differential operator 
${\cal P}$ is defined as usual,
\beq \label{zeta}
\zeta_{{\cal P}_\gamma}(s) = \sum_\lambda \lambda^{-s} \, ,
\eeq
where the sum is over eigenvalues of ${\cal P}$, and we have added
a label $\gamma$ to denote the 
boundary condition explicitly. Knowledge of the $\zeta$ function
allows of course to compute the determinant, since   $\zeta_{\cal P}'(0) = -\log\det {\cal P}$.
The basic observation is that we can write
\beq \label{bobservation}
\zeta_{{\cal P}_{\gamma_1}}(s) - \zeta_{{\cal P}_{\gamma_2}}(s) = \frac{1}{2 \pi i} 
\int d\lambda \, \lambda^{-s} \frac{d}{d\lambda} \log 
\frac{\gamma_1 p_\lambda(a) + p_\lambda'(a)}{\gamma_2 
p_\lambda(a) +  p_\lambda'(a)} \, .
\eeq
Indeed  $\lambda$ is an eigenvalue of ${\cal P}_\gamma$
if and only if
 the function $\gamma p_{\lambda} (a) + p'(\lambda) $ has
a zero; correspondingly its
logarithmic derivative  $\frac{d}{d\lambda} \log [\gamma p_{\lambda} (a) + p_\lambda'(a)] $
has a pole with unit residue and the contour integration in the complex
$\lambda$ plane reproduces the definition (\ref{zeta}).
The contour  of integration is shown in Fig. \ref{fig:roots}. The branch cut 
for $\lambda^{-s}$ is placed at an angle $\theta$ from the 
positive real $\lambda$ axis.  As it stands, this definition is 
meaningless at $s=0$.  The goal is to deform the contour to 
enclose the branch cut, and obtain an expression that extends to 
a region around $s=0$.  If the integrand behaves as 
\beq\label{eq:assumption}
\frac{d}{d\lambda} \log \frac{\gamma_1 p_\lambda(a) + 
p_\lambda'(a)}{\gamma_2 p_\lambda(a) + p_\lambda'(a)} = 
\mathcal{O}\left(\frac{1}{\lambda^{3/2}}\right)
\eeq
as $\Im \sqrt{\lambda} \rightarrow \pm \infty$, then the contour 
can be deformed and the resulting integral is well defined for 
$-1/2 < s < 1$. 
Differentiating (\ref{bobservation}) with respect to $s$ and setting
$s=0$ gives
exactly (\ref{eq:finalcoleman}), completing the proof. (See \cite{kirsten04b} for a careful
analysis of the case (\ref{eq:colemantrick})). The assumption
(\ref{eq:assumption}) can be checked for any particular case; 
we will show below that it holds in our case.\footnote{In \cite{kirsten04b}, the analogous asymptotic
behavior for the case (\ref{eq:colemantrick}) was argued to hold for general
Sturm-Liouville operators ${\cal Q}$ and ${\cal P}$.
In our case, although ${\cal P}$ is not of Sturm-Liouville
type, we will be able to check  (\ref{eq:assumption}) explicitly.}

\newsubsection{Relating the partition functions}

We are now in a position to verify equation (\ref{eq:detratio}).
We apply the  formula (\ref{eq:finalcoleman}) with ${\cal P} =``( 
- \Box + m^2)"$ the radial differential operator. The left 
extremum is taken $r=a= \epsilon$, while the right extremum is 
taken to correspond to the ``deep interior'' of AdS -- for 
example $x_0 = \infty$ in Poincar\'e coordinates and $\rho = 2$ 
in hyperbolic coordinates. We immediately have \be 
\label{relating} \frac{\det_{\gamma_1} ( - \Box + 
m^2)}{\det_{\gamma_2} (-\Box + m^2)} = \frac{\gamma_1 p(\epsilon) 
+  \epsilon p'(\epsilon)}{\gamma_2 p(\epsilon) +  \epsilon 
p'(\epsilon)}  \, , \ee where $p$ is the solution of the radial 
wave equation obeying Dirichlet boundary conditions in the 
interior of AdS. For example in Poincar\'e coordinates, $p(k, 
x_0) = \psi(k, x_0)$ as defined in (\ref{psi}). Using the 
identity \be \frac{p(\epsilon)}{\gamma p(\epsilon) +  \epsilon 
p'(\epsilon)} = Q_f  \, , \ee which was shown to hold  in section 
3 (see (\ref{eq:cftcorrelator}) and below\footnote{The argument 
below (\ref{eq:cftcorrelator}) was phrased for definiteness in 
Poincar\'e coordinates, but the conclusion is clearly general.}), 
we have finally 
\be \frac{\det_{\gamma_1} ( - \Box + 
m^2)}{\det_{\gamma_2} (-\Box + m^2)} = \frac{Q_{f_2}}{Q_{f_1}} 
\,. \ee 
It follows that the change in the partition function of 
the AdS theory is identical to that of the CFT, term by term in a 
sum over boundary modes.

\newsubsection{Explicit Analysis: Poincar\'{e} Coordinates}

To clarify the discussion above in generic coordinates, we will 
demonstrate the construction explicitly in Poincar\'{e} 
coordinates,
\beq
ds^2 = \frac{1}{x_0^2}(dx_0^2 + dx^2)\,.
\eeq
The change in the CFT effective potential from equation 
(\ref{eq:cftans}) is
\beq\label{eq:cftpoincare}
\Delta V^{CFT} = -\frac{1}{2} \int \frac{d^dk}{(2\pi)^d}\log( 1 + 
f g_k )
\eeq
where $g_k$ is the eigenvalue of the CFT propagator.  Using
\begin{eqnarray}
 \langle \bigO_{\Delta}(x)\bigO_{\Delta}(y) \rangle &=&
 \frac{1}{|x-y|^{2\Delta}} \\
 &=& \int\frac{d^dk}{(2\pi)^d} e^{-ik(x-y)} g_k
\end{eqnarray}
we find
\beq
g_k = k^{-2\nu} \pi^{d/2} 
\frac{2^{2\nu}}{\nu}\frac{\Gamma(1+\nu)}{\Gamma(d/2 -\nu)}\,.
\eeq

The change in the AdS effective potential as we switch from 
irregular to regular modes is given by equation 
(\ref{eq:deltaVads}) with $\gamma_1 \rightarrow \infty$, 
$\gamma_2 = -\Delta_-$.  

To evaluate the ratio of determinants using eq. 
\ref{eq:finalcoleman}, we must first analyze the large $\nu$ 
behavior of 
\beq
\frac{ \gamma_1 q_\lambda(a) + \epsilon q_\lambda'(a) }{ \gamma_2 
q_\lambda(a) + \epsilon q_\lambda'(a) }
\eeq
with $q_\lambda(y) = y^{d/2}\besselK_\nu(ky)$. (In this case, the 
roots in the integrand of the zeta function lie on the imaginary 
axis, so we must check that the large $\Re \nu$ behavior is no 
worse than $\bigO(\lambda^{-3/2})$.)   Using the fact that 
$\besselK_{\nu+1}(x)/\besselK_{\nu}(x) \rightarrow 2\nu/x$ as 
$\nu \rightarrow \infty$, we find
\beq
\frac{d}{d\lambda}\log\frac{ \gamma_1 q_\lambda(a) + \epsilon 
q_\lambda'(a) }{ \gamma_2 q_\lambda(a) + \epsilon q_\lambda'(a) } 
= 
\frac{a(\gamma_1-\gamma_2)}{\epsilon}\left(\frac{1}{\nu^2}\right) 
+ \bigO\left(\frac{1}{\nu^3}\right) 
\eeq 
This confirms assumption (\ref{eq:assumption}).  Finally, 
plugging into eq. (\ref{eq:finalcoleman}) and substituting $f$ 
for $\gamma$, we find precisely the answer we found for $\Delta 
V^{CFT}$ in equation (\ref{eq:cftpoincare}). This completes the 
demonstration in Poincar\,'e coordinates that
\beq
\Delta V^{AdS} = \Delta V^{CFT} \,.
\eeq

\newsubsection{Explicit analysis: Hyperbolic Coordinates}

To make contact with the results of 
\cite{gubser02} and \cite{gubser02b}, we now 
perform an explicit analysis in   hyperbolic coordinates, where the 
boundary of AdS is a sphere.   The metric is
\beq \label{hyperbolic}
ds^2 = \frac{d \rho^2}{\rho^2} + \left(\frac{4 - \rho^2}{4 
\rho}\right)^2 d\Omega_d^2 \,.
\eeq
Expanding in spherical harmonics
\beq
\phi(\rho,\Omega) = \sum_{l,m}Y_{lm}(\Omega)\phi_l(\rho)\,,
\eeq
the radial wave equation
reads
\beq
\left[ \rho^2 \frac{\partial}{ \partial \rho^2} + \rho 
\frac{\partial}{\partial \rho} + d \rho \, \frac{\rho^2 + 
4}{\rho^2 - 4} \frac{\partial}{\partial \rho} - l (l+ d -1) 
\left(\frac{4 \rho}{4 - \rho^2}  \right)^2 \right] \phi_l (\rho) 
= 0 \,.
\eeq
The unique solution regular in the ``deep interior '' at $\rho =2$ is
 \beq \label{pl}
p_l(\rho) = (\frac{4 - \rho^2}{4 \rho})^l F\left(l+\Delta_+, l+\Delta_-,
l+\frac{d}{2}+\frac{1}{2}; -\frac{(\rho - 2)^2}{8 \rho}\right)\,.
\eeq
Notice that $p_l$ is zero at $\rho=2$ together
with its first $l-1$ derivatives. The
boundary conditions at  the right extremum $b=2$
are taken to be
\beq
p_l(b) = 0 \, , \quad \frac{d^l p_l(b)}{d \rho^l}=\frac{ l!}{(-2)^l} \, .
\eeq
This is an inessential modification of the condition
(\ref{qp}).

Expanding near the boundary,
\beq
p_l(\epsilon) \approx a_l  \, \epsilon^{\Delta_+} + b_l \epsilon^{\Delta_-}  \, ,
\eeq
where
\beq
a_l = 
\frac{2^{l-1+\Delta_-}\Gamma\left(l+\frac{d}{2}+\frac{1}{2}\right) 
\Gamma\left(\Delta_--\frac{d}{2}\right)}{\sqrt{\pi}\Gamma(l+\Delta_-)} 
\,, \hspace{1cm}b_l = 
\frac{2^{l-1+\Delta_+}\Gamma\left(l+\frac{d}{2}+\frac{1}{2}\right) 
\Gamma\left(\Delta_+-\frac{d}{2}\right)}{\sqrt{\pi}\Gamma(l+\Delta_+)} 
\,.
\eeq
It is then straightforward  to check that the ratio of determinants 
\beq
\frac{\det_{\gamma_1} ( - \Box + m^2)}{\det_{\gamma_2} (-\Box + m^2)}
= \frac{\gamma_1 p_l(\epsilon) +  \epsilon p_l'(\epsilon)}{\gamma_2 p_l(\epsilon) +  \epsilon p_l'(\epsilon)} 
\eeq
agrees exactly with the expected CFT answer on ${\bf S}^d$
\beq
\frac{1 + f_1 g_l}{1 + f_2 g_l} \, ,
\eeq
where $g_l$ is the eigenvalue of the boundary Laplacian, 
(\ref{eq:gl}). To be precise, we find the CFT answer on a 
$d$-sphere with $R=1/\epsilon$, which is as expected since in 
writing the hyperbolic coordinates (\ref{hyperbolic}) we have 
normalized the boundary metric as $d \Omega_d^2/\epsilon^2$.

We  have also performed numerical checks of our method for 
computing determinant ratios, in both Poincar\'{e} and hyperbolic 
coordinates . In hyperbolic coordinates, the  boundary condition 
$\phi_l(\rho=2) = 0$
 selects (\ref{pl})
 rather than the other solution to the hypergeometric
equation, which blows up at $\rho =2$.  The  boundary
condition $\gamma \phi_l(\epsilon) + \epsilon \phi_l'(\epsilon) = 0$ then quantizes
$\nu$, giving a discrete set of eigenvalues that are multiplied
together to compute the determinant. The determinants are infinite
but their ratios are finite, converging after a few
thousand eigenvalues.

\begin{table}[t]
  \centering
  \begin{tabular}{|c|c|c|}
    \hline 
     & $R_{\rm{numerical}}$ & $R_{\rm{analytical}}$ \\
    \hline
    Point A: $\gamma_1=1\,, \gamma_2=10$ $\; \;$ & -0.0336349 & -0.0336350 \\
    \hline
    Point B: $\gamma_1=10\,, \gamma_2=100$ & 0.0882113 & 0.0882118 \\
    \hline
  \end{tabular}
  \caption{\textit{Comparison of numerical and analytical computations of (\ref{eq:detnumer}).  The analytical results were computed using eq. (\ref{eq:finalcoleman}).  The numerical results were found by computing approximately 30000 eigenvalues for each operator and extrapolating to 
$n\rightarrow \infty$.  In all cases, $d=4$, $l=3$, $\epsilon=10^{-4}$, and $m^2=-3.5$.}}\label{table:numerical}
\end{table}

\begin{figure}[h]
\centering \epsfig{file=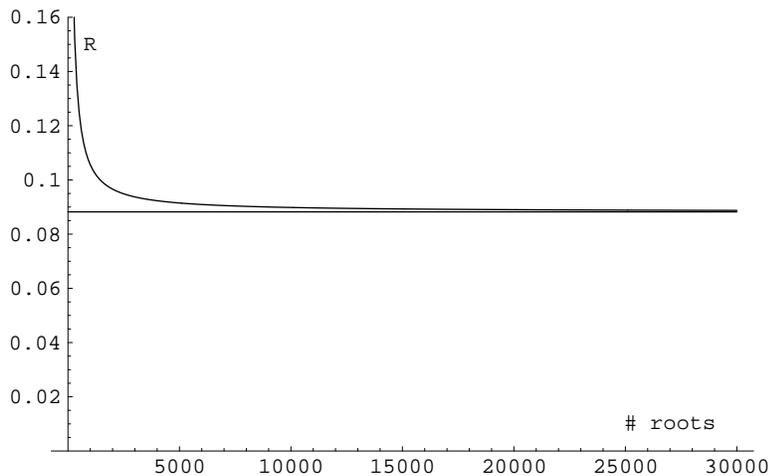, width=4in} 
\caption{\textit{Numerical value of $R$ in eq. 
(\ref{eq:detnumer}) versus the number of roots included in the 
computation.  The curve converges nicely to the horizontal line 
(the analytical answer). \label{fig:mathematica2}}}
\end{figure}

As an example, we compare the numerical and analytical results 
for two different choices of $\gamma_1, \gamma_2$ in Table 
\ref{table:numerical}.  Figure \ref{fig:mathematica2} shows a 
Mathematica plot of the convergence of 
\beq\label{eq:detnumer}
R \equiv \frac{ \det_{\gamma_1}(-\Box + m^2) } 
{\det_{\gamma_2}(-\Box + m^2)}
\eeq
for Point B.  The horizontal axis is the number of eigenvalues $n$ 
included in the computation, and the flat line is the analytical 
answer obtained from the boundary value trick.  The numerical 
solution approaches the analytical solution like $1/n$.  For each 
set of parameters, the numerical solution in Table 
\ref{table:numerical} was obtained by fitting the curve to a 
power law and extrapolating  to $n \rightarrow \infty$. 

\break

\newsection{$\Delta_-$ boundary conditions and  1PI diagrams  \label{s:legendre}}

The auxiliary field trick was used
in \cite{gubser02b} 
to show  that in the large $N$ boundary theory,
the generating functional $W_{IR}[\tilde J]$
 at the IR fixed point is  the Legendre transform of the generating
functional $W_{UV}[ J]$
at the UV fixed point (see \cite{Elitzur:2005kz} for a recent discussion). 
This agrees with 
the recipe postulated by Klebanov
and Witten \cite{witten99} for the evaluation of AdS/CFT correlators
with $\Delta_-$ boundary conditions: namely one is instructed to
first evaluate the correlators with $\Delta_+$ boundary conditions,
using the standard algorithm, and then to perform a Legendre transform
with respect to the source $\tilde J$ for the operator ${\cal O}_{\Delta_+}$.
While this is  a consistent state of affairs,
it would be more satisfactory
to treat the two boundary conditions on a more symmetric footing,
and have an intrinsic algorithm that directly computes 
 AdS/CFT correlators with $\Delta_-$ boundary conditions.

Actually,  we have already stated such an algorithm  in section 
3: evaluate the on-shell bulk action as a functional of the 
boundary source $\phi_b \equiv  J$ of eq. (\ref{boundaryvalue}), 
imposing the  mixed boundary condition with $\gamma = - 
\Delta_-$. In section 3 we were dealing with two-point functions, 
and thus restricting to the quadratic part of the action. For 
higher point function, we  proceed in the standard way 
\cite{gub98, wit98}, treating the interactions perturbatively, 
and computing AdS/CFT correlators as sums of diagrams built with 
bulk-to-boundary propagators, bulk-to-bulk propagators and bulk 
vertices (see {\it e.g.} \cite{exchange}). Our algorithm is 
simply  stated: use the propagators appropriate to $\Delta_-$ 
boundary conditions, and otherwise proceed as usual.

Let us take a closer look at the propagators, contrasting  
regular and irregular boundary conditions. It will be convenient 
to work  in momentum space. The bulk-to-boundary propagators for 
$\Delta_+$ and $\Delta_-$ are given by (\ref{P}), respectively 
with $\gamma = \infty$ and $\gamma = -\Delta_-$, \bea 
P_{\Delta_+}(k) &  = &\frac{\psi(k, x_0) }{ \psi(k, \epsilon) }  
\stackrel{\epsilon \to 0} {\longrightarrow }  
  \frac{2^{1-\nu}}{\Gamma(\nu)} (\epsilon  k)^\nu \,   x_0^\nu  {\cal K}_\nu(k x_0) \, ,\\
P_{\Delta_-}(k) &  = &\frac{\psi(k, x_0)}{- \Delta_- \psi(\epsilon, k) + \epsilon \, \partial \psi(k , \epsilon)  } 
 \stackrel{\epsilon \to 0} {\longrightarrow }    \frac{2^{1+\nu}}{ 2 \nu \Gamma(-\nu)} (\epsilon  k)^{-\nu} \,   x_0^\nu  {\cal K}_\nu(k x_0),
\eea
where 
we  have reabsorbed  the
prefactor of $1/\gamma \to 0$ in $P_{\Delta_+}$ in a  redefinition of the source.
The two propagators have the same functional
form in $x_0$. This is bound to happen since there
is a unique solution of the wave equation regular in the interior.
 They differ however by an important $k$ dependent factor,
\be
\frac{P_{\Delta_+}(k)} {P_{\Delta_-}(k)}   \sim k^{2 \nu} \, .
\ee
The bulk-to-bulk  propagators are
\bea
G_{\Delta_+}( k; x_0, y_0)  & = & (x_0y_0)^{d/2}I_\nu(k z_0^<)\bK_\nu(kz_0^>) \\
G_{\Delta_-}( k; x_0, y_0)  & = & 
(x_0y_0)^{d/2}I_{-\nu}(kz_0^<)\bK_\nu(kz_0^>) \, , \eea where 
$z_0^<$ is the smaller of the radial coordinates $x_0$, $y_0$ and 
$z_0^>$ is the larger.\footnote{For simplicity, we have written 
the expressions of the bulk-to-bulk propagators in the limit 
$\epsilon \to 0$. This is legitimate in most cases. For special 
correlators (for example extremal correlators \cite{extremal}), 
if one is interested in their exact normalization,  it may be 
necessary to use the more cumbersome expressions with finite 
$\epsilon$ and take the limit $\epsilon \to 0$ at the end of the 
calculation.} We observe that their difference takes the 
factorized form \be  \label{diff} G_{\Delta_+} (k, x_0, 
y_0)-G_{\Delta_-}(k, x_0, y_0) =  -\frac{2 \sin(\pi \nu)}{\pi 
}(x_0 y_0)^{d/2}\bK_\nu(k x_0) \bK_\nu(k y_0) \,. \ee

It is not immediately obvious
that exchanging $\Delta_+$ with $\Delta_-$ propagators
 is equivalent to the Legendre transform
recipe stated above. We are now going to show that this
is the case. We will briefly illustrate
our claim  for 4pt functions -- it is easy to
fill in the details and 
 generalize the argument to arbitrary $n$-point functions.
As is familiar, the Legendre transform of the generator of connected
correlation functions is the generator of one particle
irreducible (1PI) correlation functions, with  external legs amputated.
We use the following diagrammatic
notation for the  action of the Legendre transform 
 on a 4pt function,\footnote{Of course, we are  taking the Legendre transform
of a generating functional, not of a single correlator.  What we
mean by the Legendre transform of a 4pt function is actually
the fourth functional derivative of the Legendre-transformed
generating functional.}
\begin{equation}\label{eq:legidentity}
L\left[\myg{-20pt}{fullfour.epsi}\right] =
\myg{-20pt}{pifour.epsi}
\end{equation}
Black dots represent connected correlators and
hatched dots represent 1PI correlators. The connected 4-point function can be
expanded with 1PI vertices and full propagators,
\begin{equation}
\myg{-20pt}{fullfour.epsi} = \myg{-30pt}{fourwlegs.epsi} +
\myg{-50pt}{fourexchange.epsi} + \mbox{crossed diagrams}\,.
\end{equation}
Therefore we can take the Legendre transform of a correlator by
subtracting reducible parts and removing the external legs,
\begin{equation}\label{eq:legprocess}
L\left[\myg{-20pt}{fullfour.epsi}\right] =
\left(\myg{-20pt}{fullfour.epsi} -
\myg{-50pt}{fourexchange.epsi}-\mbox{crossed }\right)\left(\myg{-5pt}{leg.epsi}\right)^{-4}\,.
\end{equation}
On the other hand, AdS diagrams are drawn in the usual way, with  a circle on the outside
that represents the boundary. The statement of the AdS/CFT
correspondence is 
\begin{equation}
\myg{-30pt}{fullfourlab.epsi} = \myg{-30pt}{exchangereg.epsi} +
\mbox{crossed diagrams} + \cdots
\end{equation}
The dots stand for diagrams without explicit dependence on the
field dual to $\bigO_{\Delta_+}$.  Here it is understood that the 
AdS diagrams on the rhs are built with  propagators obeying the 
``regular'' (Dirichlet) boundary conditions. We wish to show that 
taking the Legendre transform is equivalent to using instead the 
``irregular'' propagators. The external operators ${\cal 
O}_{\Delta_i}$, $i=,1 \dots, 4$ need not be the same as ${\cal 
O}_{\Delta_+}$. Let us first assume for simplicity that they are 
all different from ${\cal O}_{\Delta_+}$. We are then instructed 
to take the Legendre transform with respect to the source of 
$\bigO_{\Delta_+}$, leaving the operators $\bigO_{\Delta_i}$ 
unchanged.  The identity (\ref{eq:legprocess}), where external 
legs are removed on the right-hand side, assumed that only a 
single operator was involved so all of the external legs belonged 
to that operator. In the present case, the proper prescription is 
to leave the external legs intact. Diagrammatically, the 
statement is
\begin{equation}
L\left[\myg{-30pt}{fullfourlab.epsi}\right] =
\myg{-30pt}{fullfourlab.epsi} - \myg{-55pt}{exchangelab.epsi} -
\mbox{crossed}
\end{equation}
In the second diagram on the rhs, the internal wavy line with a 
small black dot represents the $\bigO_{\Delta_+}$ 2pt correlator, 
$\langle \bigO_{\Delta_+} (k) \bigO_{\Delta_+}(k)\rangle \sim 
k^{2 \nu} $, while the external solid lines with little black 
dots are the 2pt correlators of the $\bigO_{\Delta_i}$ operators. 
In momentum space the second diagram can be also be written 
\begin{equation}
\left(\myg{-40pt}{threepoint1.epsi}\right)\left(\myg{-40pt}{threepoint2.epsi}\right)\left(\myg{-3pt}{wavyprop.epsi}\right)^{-1} \, .
\end{equation}
 Translated into AdS
language, this says (we are being a little schematic and dropping 
constant factors)
\begin{samepage}
\begin{eqnarray}\label{eq:finalleg}
L\left[\myg{-30pt}{exchangereg.epsi}\right] &=&
\myg{-30pt}{exchangereg.epsi}\\-& k^{-2\nu}&
\left(\myg{-20pt}{threepointads1.epsi}\myg{-20pt}{threepointads2.epsi}\right)\,.\nonumber
\end{eqnarray}
\end{samepage}
We need to show that
the right-hand side of
(\ref{eq:finalleg}) is actually equal to the exchange diagram with
a bulk $\Delta_-$ propagator, so that
\begin{equation} \label{dlegendre}
L\left[\myg{-30pt}{exchangereg.epsi}\right] =
\myg{-30pt}{exchangeirreg.epsi}\, .
\end{equation}
Indeed, let us evaluate
the difference of exchange diagrams with regular and irregular
boundary conditions,
\vspace{1cm}
\begin{equation}
\myg{-30pt}{adsreglab.epsi} - \myg{-30pt}{adsirreglab.epsi} \, ,
\end{equation}
whose analytic expression  is 
\begin{eqnarray}
\int \frac{dz_0}{z_0^{d+1}} \int \frac{dw_0}{w_0^{d+1}}
P_{\Delta_1}(\vec{p}_1, z_0) P_{\Delta_2}(\vec{p}_2, z_0)
P_{\Delta_3}(\vec{p}_3, w_0)
P_{\Delta_4}(\vec{p}_4,w_0)\nonumber\\
\hspace{1cm}\times \big[G_{\Delta_+}(\vec{k}, z_0,
w_0)-G_{\Delta_-}(\vec{k}, z_0, w_0)\big]\,.
\end{eqnarray}
 Using (\ref{diff}), the integrand is proportional to
\begin{equation}
P_{\Delta_1}(\vec{p}_1, z_0) P_{\Delta_2}(\vec{p}_2, z_0)
P_{\Delta_3}(\vec{p}_3, w_0) P_{\Delta_4}(\vec{p}_4,w_0)
\big[z_0^{d/2}\bK_\nu(kz_0)w_0^{d/2}\bK_\nu(kw_0)\big]\,.
\end{equation}
 The difference between the bulk propagators (the term
in brackets) is seen to be identical to product of two bulk-to-boundary
propagators  $P_{\Delta_+}$, missing the factors of $k^{\nu}$, so we have found
\begin{samepage}\begin{eqnarray} \myg{-30pt}{exchangereg.epsi} -
\myg{-30pt}{exchangeirreg.epsi} =
\hspace{2cm}\nonumber\\
 k^{-2\nu}
\left(\myg{-30pt}{threepointads1.epsi}
\myg{-30pt}{threepointads2.epsi}\right)
\end{eqnarray}\end{samepage}
Comparing to (\ref{eq:finalleg}), this completes the
proof of (\ref{dlegendre}).

Finally, let us consider the case where some of the external operators ${\cal O}_{\Delta_i}$
are  also  equal to ${\cal O}_{\Delta_+}$.  In taking the Legendre transform, we
are now further instructed to  amputate each external leg involving
a ${\cal O}_{\Delta_+}$, see (\ref{eq:legprocess}). In momentum space,
this amounts to removing  factors of $\langle {\cal O}_{\Delta_+}  (k_i)  {\cal O}_{\Delta_+}(-k_i) \rangle \sim k_i^{-2 \nu}$
for each external momentum $k_i$ corresponding to a boundary insertion of ${\cal O}_{\Delta_+}$.
But this is exactly equivalent to changing boundary conditions
for  the bulk-to-boundary propagators, $P_{\Delta_+}(k_i) \to P_{\Delta_-}(k_i)$!
This concludes the argument that correlation
functions computed with $\Delta_-$ propagators
are precisely the Legendre transform of correlations
functions computed with $\Delta_+$ propagators.

\vspace{1cm}

 {\bf Acknowledgments}

 \medskip

 It is a pleasure to thank Igor Klebanov for useful discussions
 at various stages of this project, and Dan Freedman and Igor Klebanov
 for critical reading of the manuscript. 
 This material is based in part upon work supported by the National Science Foundation Grant No. PHY-0243680. Any opinions, findings, and
conclusions or recommendations expressed in this material are
those of the authors and do not necessarily reflect the views of
the National Science Foundation.

\medskip

\appendix

\newsection{One--loop effective potential in AdS}

The change in the central charge of the bulk theory was computed
by Gubser and Mitra \cite{gubser02}.  We will review the
computation here, using a different method to regulate the
infinities but otherwise following their calculation.

We are interested in  the one-loop 
 AdS effective action,\footnote{In this appendix we adopt Lorentzian signature.}
\beq\label{eq:defveff}
W = -\frac{i}{2} {\rm Tr} \log(-\Box + m^2)\,.
\eeq
Taking the derivative of this expression,
\beq
\frac{\partial W}{\partial m^2} =
\frac{i}{2}{\rm Tr} \left(\frac{1}{-\Box+m^2}\right)\,.
\eeq
Therefore the logarithm in the effective action can be written
as an integral,
\beq
W = -\frac{i}{2} {\rm Tr} \int dm^2 G
\eeq
where $G$ is the scalar propagator, defined as the inverse of the
wave operator.  The trace is an integral  $\int d^{d+1}x  \sqrt{g}$ 
over the AdS bulk. 
The maximal symmetry of AdS implies that $G(x,x)$ is independent of $x$,
so that the trace   contributes only the overall AdS volume  ${\rm Vol}_{AdS}$.
We can then define the effective potential $V(m^2)$ as
\beq
V(m^2) \equiv \frac{ W(m^2)}{{\rm Vol}(AdS)}=  -\frac{i}{2}\int_{m^2}^{\infty} d\tilde{m}^2 G(x,x;
\tilde{m}^2)\,.
\eeq
A more rigorous derivation using the
DeWitt-Schwinger representation of the propagator is given in
\cite{birrel82}, Section 6.1.

Generally, both the integrand and the integral over $\tilde{m}^2$
are divergent.  Since we are only interested in the
\textit{change} in the effective potential, we can regulate the UV
divergence in the integrand by computing the difference in $V$
going from regular ($+$) to irregular ($-$) boundary conditions,
\beq\label{eq:veff}
V_+ - V_- = -\frac{i}{2} \int_{m^2}^\infty
d\tilde{m}^2\left(G_{\tilde{\Delta}_+}(x,x) -
G_{\tilde{\Delta}_-}(x,x)\right)\,.
\eeq
The integral, however, still diverges. To regulate it, the authors
of \cite{gubser02} split the integral into two parts,
\begin{equation}\label{eq:gubsercas}
V_+(m^2)-V_-(m^2) = \frac{i}{2}\int_{m_{BF}^2}^{m^2} d\tilde{m}^2
\big(G_{\tilde\Delta_+}(x,x)-G_{\tilde\Delta_-}(x,x)\big) +
\big[V_+(m_{BF}^2)-V_-(m_{BF}^2)\big]\,.
\end{equation}
The integral is finite.  The second term is the change in the
effective potential for a field with mass saturating the
Breitenlohner-Freedman bound  $m_{BF}^2 L^2 =
-d^2/4$. This is the lowest mass for which a scalar field can be
quantized consistently on AdS, and corresponds to a boundary
operator with scaling dimension $\Delta = d/2$.

It seems reasonable to expect that the change in the effective
potential to vanish at the BF bound, leaving only the finite terms
in equation (\ref{eq:gubsercas}). A heuristic argument
is as follows \cite{gubser02}. The expression for
the mode-sum vacuum energy in the Hamiltonian formalism 
is
\begin{equation}
E_{vac} = \frac{1}{2}\sum_k \omega_k\,.
\end{equation}
For a scalar field, the mode frequencies in global
AdS coordinates are $\omega = l + 2n +
\Delta_\pm$ (see {\it e.g. }\cite{balasu98}).  Therefore, at the BF bound where
$\Delta_+ = \Delta_-$, regular and irregular modes contribute
equally to the vacuum energy.  Thus we have $E_{vac}^+(m_{BF}^2)
- E_{vac}^-(m_{BF}^2) = 0$.  However, there is a potential loophole
in this argument: 
the quantity $E_{vac}^+ -
E_{vac}^-$ 
cannot
necessarily be identified with the change in the effective
potential $V_+ - V_-$. These two quantities are
equal in flat space, but they can be different in curved space
 if the $g_{tt}$ component of the metric
is non-trivial (see {\it e.g. }\cite{camporesi92}). They
 {\it are} in fact different  in our case,
as can be verified by performing the sum explicitly with an exponential or zeta-function
regulator.

Therefore, we will use a different method that allows us to
compute $V$ directly from (\ref{eq:veff}). We work in global
coordinates,
\be
ds^2 = L^2 \left[ - \sec^2 \rho\, dt^2 + \sec^2 \rho\, d\rho^2 + \tan^2 \rho\, d\Omega^2_{d-1}\right] \,.
\ee
We begin by writing the scalar
propagator as a mode-sum,
\begin{equation}\label{eq:propmodesum}
i G_{\Delta_\pm}(x,x') =
\theta(x,x')\sum_{n,l,m}\Psi_{nlm}^{\Delta_\pm^*}(x)\Psi_{nlm}^{\Delta_\pm}(x')
+\theta(x',x)\sum_{n,l,m}\Psi_{nlm}^{\Delta_\pm^*}(x')\Psi_{nlm}^{\Delta_\pm}(x)\,,
\end{equation}
where the sum is over all normalized modes of the scalar field and
$\theta(x,x')$ is the AdS time-ordering operator. Note that the
modes are taken on-shell, $\omega = l + 2n + \Delta_\pm$.  This is
easily seen to be a valid propagator by applying the wave operator
$-\Box + m^2$. Acting on the $\theta$-functions gives a delta
function $\delta(x-x')$, and the equation of motion for
$\Psi_{nlm}^{\Delta_\pm}$ gives zero for $x\neq x'$.

Plugging in the scalar modes (from {\it e.g.} \cite{balasu98}), we have
\begin{eqnarray}
i G_\Delta(x,x') &=& \sum_{n,l,m}N_\Delta^2
e^{-i\omega(t-t')}Y_{lm}(\Omega)Y_{lm}(\Omega')
(\sin\rho)^l(\cos\rho)^\Delta(\sin\rho')^l(\cos\rho')^\Delta\nonumber\\
& & \times P_n^{(l+d/2-1, \Delta-d/2)}(\cos 2\rho)P_n^{(l+d/2-1,
\Delta-d/2)}(\cos 2\rho')\,,
\end{eqnarray}
where $N_\Delta^2$ is a normalization constant that can be
computed from the scalar inner product and $P_n^{a,b}$ is a Jacobi
polynomial. The maximal symmetry of AdS allows us translate to $x'
= 0$, considerably simplifying the sum by ensuring that only
$l=0\,,\, m=0$ modes will contribute. All other terms are zero by
virtue of the $(\sin\rho')^l$ factor. $Y_{00}(\Omega)$ is constant
and normalized on the sphere $S^{d-1}$, so the propagator is
\begin{equation}
i G_\Delta(x) =
\frac{(\cos\rho)^\Delta}{{\rm Vol}(S^{d-1})}\sum_{n=0}^\infty N_\Delta^2
P_n^{(d/2-1, \Delta-d/2)}(1)P_n^{(d/2-1, \Delta-d/2)}(\cos
2\rho)e^{-i (2n+\Delta)t}\,.
\end{equation}
This can be summed to give the usual expression for the propagator
in global coordinates. To regulate the coincidence limit of
(\ref{eq:propmodesum}), which does not depend on $x$, we set
$\rho=\rho'$ and $\Omega=\Omega'$ but maintain $t\neq t'$ until
the very end.   After interchanging the order of the limit, sum,
and integral, this gives
\begin{equation}
V(m^2) = - \lim_{t\rightarrow 0} \frac{1}{4 L^d
\pi^{d/2}\Gamma(\frac{d}{2})} \sum_{n=0}^\infty
\int_{\nu^2}^\infty d\tilde{\nu}^2
\frac{\Gamma(n+\frac{d}{2})\Gamma(n+\frac{d}{2}+\tilde\nu)}{\Gamma(n+1)\Gamma(n+1+\tilde\nu
)}e^{-i (2n+\frac{d}{2}+\tilde\nu)t}\,.
\end{equation}
Everything but the final limit is finite.  To regulate the
$t\rightarrow 0$ divergence, we expand the answer as a power
series in $t$ and take the difference $V_+ - V_-$ before taking
the limit.  The result is finite.  

For even $d$, the integral is simple and we obtain the following:
\begin{itemize}\label{eq:adsanswer}
    \item $\mathbf{d=2}:$\hspace{1.5cm}  $V_+ - V_- =
    -\frac{\nu^3}{6\pi L^2}$
    \item $\mathbf{d=4}:$\hspace{1.5cm} $V_+ - V_- =
    \frac{1}{12\pi^2 L^4}\left(-\frac{\nu^3}{3}+\frac{\nu^5}{5}\right)$
    \item $\mathbf{d=6}:$\hspace{1.5cm} $V_+ - V_- = \frac{1}{120 \pi^3 L^6}\left(-\frac{4}{3} \nu^3+\nu^5-\frac{\nu^7}{7}\right)$
    \item $\mathbf{d=8}:$\hspace{1.5cm} $V_+ - V_- = \frac{1}{1680 \pi^4 L^8}\left(-12\nu^3+\frac{49}{5}\nu^5 - 2 \nu^7 + \frac{\nu^9}{9}\right)$
\end{itemize}
These results agree with  \cite{gubser02},
thereby confirming their a priori assumption that $ V_+ (m_{BF}^2)= V_-(m_{BF}^2)$.
 They are
also compatible with the
calculation in  \cite{caldarelli98} 
of  the (Euclidean) one-loop effective potential by zeta-function regularization,
if one assumes that we can obtain $V_-$ by simply replacing
$\Delta_+\rightarrow \Delta_-$ in the final expressions for $V_+$ quoted in  \cite{caldarelli98}.

\begingroup\raggedright

\providecommand{\href}[2]{#2}

\endgroup

\end{document}